\documentclass[noshowpacs,amsmath,
twocolumn,
superscriptaddress,
8pt,
aps,prb]{revtex4-1}
\usepackage{setspace}
\usepackage{amsmath}
\usepackage{graphicx}
\usepackage[nearskip,margin = 0pt]{subfig}

\usepackage{verbatim}
\usepackage{amsfonts}
\usepackage{amssymb}
\usepackage{epstopdf} 
\usepackage{xcolor}
\DeclareGraphicsExtensions{.pdf,.eps,.png,.jpg,.mps} 

\begin{document}

\title{Spatial-mode-interaction-induced dispersive-waves and their active tuning in microresonators}

\author{Qi-Fan Yang$^{\ast}$, Xu Yi$^{\ast}$, Ki Youl Yang and Kerry Vahala$^{\dagger}$\\
T. J. Watson Laboratory of Applied Physics, California Institute of Technology, Pasadena, California 91125, USA.\\
$^{\ast}$These authors contributed equally to this work.\\
$^{\dagger}$Corresponding author: vahala@caltech.edu}
\maketitle

\noindent {\bf The nonlinear propagation of optical pulses in dielectric waveguides and resonators provides a laboratory to investigate a wide range of remarkable interactions. Many of the resulting phenomena find applications in optical systems. One example is dispersive wave generation, the optical analog of Cherenkov radiation. These waves have an essential role in fiber spectral broadeners that are routinely used in spectrocopy and metrology. Dispersive waves form when a soliton pulse begins to radiate power as a result of higher-order dispersion. Recently, dispersive wave generation in microcavities has been reported by phase matching the waves to dissipative Kerr cavity (DKC) solitons. Here, it is shown that spatial mode interactions within a microcavity can also be used to induce dispersive waves. These interactions are normally avoided altogether in DKC soliton generation. The soliton self frequency shift is also shown to induce fine tuning control of the dispersive wave frequency. Both this mechanism and spatial mode interactions provide a new method to spectrally control these important waves.}

If the spectrum of a soliton pulse extends into regions where second-order dispersion changes sign, then radiation into a new pulse, the dispersive wave, may occur at a phase matching wavelength \cite{wai1986nonlinear,akhmediev1995cherenkov}. The generation of these waves is analogous to Cherenkov radiation \cite{cherenkov1934visible} and greatly extends the spectral reach of optical pulses \cite{dudley2006supercontinuum}. The recent ability to control dispersion in microresonators has allowed accurate spectral placement of dispersive waves relative to a radiating cavity soliton \cite{brasch2016photonic}. Such dispersion-engineered control has made possible 2f-3f self referencing of frequency microcombs \cite{brasch2016self} and octave-spanning double-dispersive waves \cite{li2015octave}. To date, dispersive wave generation in both resonators and in optical fiber has resulted from manipulation of geometrical dispersion in conjunction with the intrinsic material dispersion of the dielectric \cite{brasch2016photonic,dudley2006supercontinuum}. 

Here, a different mechanism for dispersive wave generation is demonstrated: spatial mode interaction within a microresonator. These mode interactions often frustrate the formation of DKC solitons \cite{herr2014mode} and, as a result, resonators are typically designed to minimize or exclude entirely the resulting modal avoided crossing \cite{herr2014temporal,brasch2016photonic,yi2015soliton,wang2016intracavity,joshi2016thermally}. Also, while dispersive-wave phase matching is normally induced by more gradual variations in dispersion, spatial mode interactions produce spectrally abrupt variations that can activate a dispersive wave in the vicinity of a narrow-band soliton. These interactions can, in principle, occur mutliple times, suggesting the possibility of dispersive wave multiplets scattered from a single soliton. Below, the demonstration of dispersive wave generation is presented after characterizing two strongly interacting spatial mode families. The phase matching of this wave is then studied including for the first time the effect of soliton frequency offset relative to the pump as is caused by soliton recoil or by the Raman-induced soliton self-frequency shift (SSFS). It is shown that this mechanism enables active tuning control of the dispersive wave by pump tuning.

\begin{figure}[!ht]
\includegraphics[width=\linewidth]{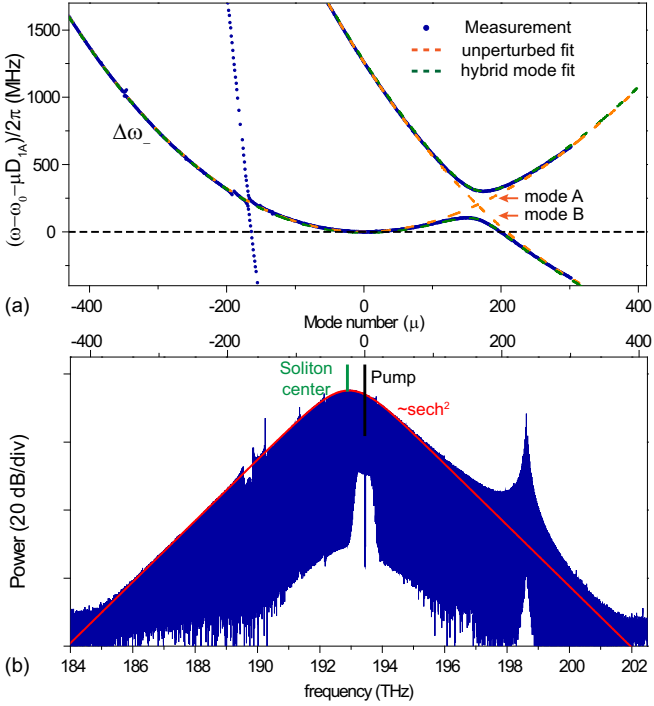}
\caption{Dispersive wave generation by spatial mode interaction. (a) Measured relative frequencies (blue points) of the soliton-forming mode family and the interaction mode family. Mode number $\mu =0$ corresponds to the pump laser frequency of 193.45 THz (1549.7 nm). Hybrid mode frequencies calculated from eqn.(1) are shown in green and the unperturbed mode  families are shown in orange. The dashed black line gives the phase matching condition in cases where the soliton repetition rate equals the microcavity FSR at $\mu = 0$. (b) Typical DKC soliton optical spectrum with dispersive wave feature. For comparison, a $Sech^2$ fitting is shown in red. The pump frequency (black) and soliton center frequency (green) indicate a Raman-induced soliton self-frequency shift.}
\label{figure1}
\end{figure}

In a resonator, the formation of dissipative Kerr cavity (DKC) solitons induces phase locking of the longitudinal modes belonging to a specific spatial mode family of the resonator \cite{matsko2011mode,herr2014temporal}. In the absence of third-order dispersion, a soliton will form by balance of the second-order dispersion term with the Kerr nonlinearity. The introduction third-order dispersion will perturb the soliton and can lead to radiation of a dispersive wave \cite{brasch2016photonic}. To form solitons in microcavities interactions between the soliton mode family and other spatial mode families are eliminated or minimized so as to ensure that second-order dispersion is dominant \cite{herr2014mode}. Here, these interactions are intentionally used to enable dispersive wave generation. 

In the experiment, an ultra-high-Q silica micro-resonator (3 mm diameter) with a 22 GHz free-spectral-range (FSR) was prepared \cite{lee2012chemically}. Typical intrinsic quality factors were in excess of 300 million (cavity linewidths were less than 1MHz). Mode dispersion was characterized from 183.92 THz (1630 nm) to  199.86 THz (1500 nm) by fiber-taper coupling a tunable external-cavity diode laser (ECDL) and calibrating the frequency scan using a Mach-Zehnder interferometer (MZI) \cite{yi2015soliton}. Multiple mode families were observed and their measured frequency dispersion spectra are presented as the blue points in fig. 1(a). In the plot, a linear dispersion term corresponding to the FSR of the soliton-forming mode family at mode number zero is subtracted so that a $relative$-$mode$-$frequency$ is plotted. Mode zero is by convention the mode that is optically pumped to form the soliton. Three weak perturbations of the soliton mode family dispersion are observed for $\mu < 0$. The mode family associated with one of the perturbations is plotted as the nearly vertical line of blue points. A much stronger interaction occurs near $\mu = 165$ causing a strong avoided mode crossing that redirects the soliton-forming branch to lower relative mode frequencies.

\begin{figure*}[!ht]
\includegraphics[width=17.5cm]{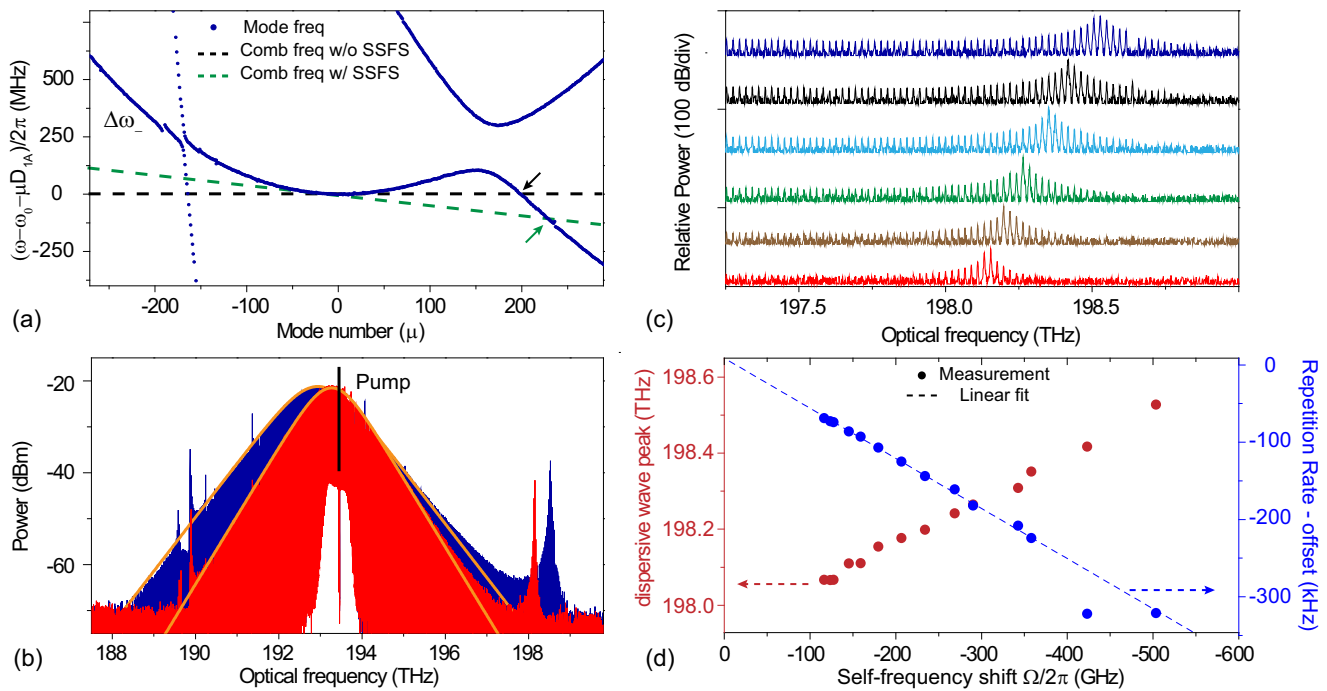}
\caption{Dispersive wave phase matching condition and Raman-induced frequency shift. (a) Soliton and interaction mode family dispersion curves are shown (see fig. 1a) with phase matching dashed lines (eqn. (4)). The black line is the case where $\omega_r = D_{1A}$ and the green line includes a Raman-induced change in $\omega_r$. The intersection of the soliton branch with these lines is the dispersive wave phase matching point (arrows). (b) Soliton optical spectra corresponding to small (red) and large (blue) cavity-laser detuning ($\delta \omega$). $Sech^2$ fitting of the spectrum envelope is shown as the orange curves. (c) High resoluton scan of dispersive wave spectra with cavity-laser detuning (soliton power and bandwidth) increasing from lower to upper trace. (d) Dispersive wave peak frequencies (orange points) and soliton repetition rate (blue points) are plotted versus soliton self-frequency shift. The linear fit (dash line) agree with the model. The offset for the repetition rate vertical scale is $=D_{1A}=21.9733$ GHz.}
\label{figure2}
\end{figure*}

The dispersion of the two interacting mode families can be accurately modeled using a coupled mode approach. Accordingly, consider two mode families ($A$ and $B$) that initially do not interact and that feature frequency dispersion spectra $\omega_{A,B} (\mu)$. An interaction between the mode families is introduced that is characterized by a coupling rate $G$.  The coupling produces two hybrid mode families with upper/lower-branch mode frequencies $\omega_{\pm} (\mu)$ given by the following expression \cite{haus1991coupled,wiersig2006formation,liu2014investigation},
\begin{equation}
\omega_{\pm} (\mu) = \frac{\omega_A (\mu)+\omega_B (\mu)}{2} \pm \sqrt[]{G^2+[\omega_A(\mu)-\omega_B(\mu)]^2/4}
\end{equation}
Note that in the limit of $|\omega_A(\mu)-\omega_B(\mu)| \gg G$, the frequencies $\omega_{\pm}$ approach the frequencies, $\omega_{A,B} (\mu)$, of the non-interacting mode families. The form of $\omega_{A,B} (\mu)$ are determined using this fact by fitting them within the regions $\mu<50$ and $\mu>280$ of the measured dispersion spectra to the following equation: $\omega_{A,B} (\mu) = \omega_{A,B}(0)+ D_{1A,B} \mu + D_{2,A,B} \mu^2 / 2 + D_{3,A,B} \mu^3 / 6 $, which is a third-order Taylor expansion of each mode family about mode number $\mu = 0$. The corresponding fits are shown as the dashed orange curves in fig. 1a. For mode family $A$: $D_{1A}/2\pi=21.9733$ GHz, $D_{2A}/2\pi=15.2$ kHz, $D_{3A}/2\pi=-14.7$ Hz; and for mode family $B$: $D_{1B}/2\pi=21.9654$ GHz, $D_{2B}/2\pi=18.6$ kHz, $D_{3B}/2\pi=-17.2$ Hz and $\omega_B(0)-\omega_A(0)=1.261$ GHz. The coupling coefficient, $G$, is determined by the minimum frequency difference of two branches and gives $G/2\pi$=106.5 MHz. Using these parameters, hybrid mode frequencies $\omega_{\pm}(\mu)$ (eqn. (1)) are plotted in green in fig. 1a and show good agreement with the measured dispersion branches. An improved fitting is possible by applying a least squares approach using eqn. (1). As an aside, the definition of the relative frequency for fig. 1a (and fig. 2a) is $\omega_0 \equiv \omega_A(0)$. 

The optical spectrum of a DKC soliton pumped at $\mu = 0$ (193.45 THz or 1549.7 nm) using a fiber laser is presented in fig. 1b. The soliton is triggered and stabilized using the method described elsewhere \cite{yi2015soliton,yi2016active}. For comparison, the ideal hyperbolic $Sech^2$ spectral profile that would occur under conditions of pure second-order dispersion \cite{herr2014temporal,yi2015soliton,brasch2016photonic} (mode A dashed orange curve in fig. 1a) is provided as the red envelope in fig. 1b. A small soliton self-frequency shift (SSFS) \cite{brasch2016photonic,karpov2016raman,yi2015soliton} is apparent in the measured soliton spectrum as indicated by the spectral displacement of the soliton spectral center relative to the pumping frequency. The perturbations to the ideal spectral envelope that are caused by both the weak modal crossings ($\mu < 0$) as well as the strong avoided modal crossing are apparent. For $\mu>0$ a dispersive wave feature (maximum near 198.62 THz or $\mu = 235$) is apparent. The phase matching condition for this dispersive wave is considered next.


Phase matching between the soliton and the dispersive wave occurs when the $\mu^{th}$ soliton line at $\omega_{p}+ \omega_r \mu$ ($\omega_p$ is the pump frequency and $\omega_r$ is the soliton repetition frequency) is resonant with the $\mu^{th}$ frequency of the soliton-forming mode family, i.e., $\omega_-(\mu) = \omega_{p}+ \omega_r \mu$ (the soliton forms on the lower frequency branch, $\omega_-(\mu)$).
As an aside, the Kerr shift for mode $\mu$ is much smaller than other terms in this analysis and is neglected in the phase matching condition. So that it is possible to use a graphical interpretation of the phase matching condition based on the relative frequency of fig. 1a, $ \omega_0 + D_{1A}\mu$ is subtracted from both sides of the phase matching condition to give the following condition,
\begin{equation}
\Delta \omega_- (\mu) = (\omega_r - D_{1A}) \mu - \delta \omega 
\end{equation}
where $\delta \omega \equiv \omega_0 - \omega_P$ is the detuning of the resonator relative to the pump frequency, and where $\Delta \omega_- (\mu) \equiv \omega_- (\mu) -\omega_0 - D_{1A}\mu$ is the soliton forming branch plotted in fig. 1a. 

If the soliton repetition frequency equals the FSR at $\mu=0$ (i.e., $\omega_r = D_{1A}$), then the r.h.s. of eqn. (2) is the horizontal dashed black line in fig. 1a (repeated in fig. 2a). Under these circumstances the dispersive wave phase matches to the soliton pulse at the crossing of that line with the soliton-forming mode branch. However, while the mode dispersion profile ($\Delta \omega_-(\mu)$) is determined entirely by the resonator geometry and the dielectric material properties, the soliton repetition rate $\omega_r$ depends upon frequency offsets between the pump and the soliton spectral maximum. Defining this offset as $\Omega$, the repetition frequency is given by the following equation \cite{matsko2013timing,jang2015temporal},
\begin{equation}
\omega_r=D_{1A}+\frac{D_{2A}}{D_{1A}} \Omega.
\end{equation} 
The offset frequency $\Omega$ can be caused by soliton recoil due to a dispersive wave and also by the Raman-induced soliton self-frequency shift (SSFS) \cite{brasch2016photonic,karpov2016raman,yi2015soliton}. In this work, $\Omega$ is dominated by the Raman interaction, because the typical dispersive wave power is $ < 0.2 \%$ of the soliton power and causes a negligible dispersive wave recoil. It is also noted that photo-thermal-induced change in $D_{1A}$ is another possible contribution that will vary $\omega_r$ as pumping is varied \cite{del2008full}. However, it is estimated to be $\sim -4.5$ kHz/mW (by measurement of resonant frequency photo-thermal shift of $\sim -40$ MHz/mW). With total soliton power less than 1 mW \cite{yi2015soliton}, this photo-thermal induced change is therefore also determined to be negligible compared with the effect of the Raman self-frequency-shift in eqn.(3). 

Combining eqns. (2) and (3) gives the following phase matching condition: 
\begin{equation}
\Delta \omega_- (\mu)= \mu \frac {D_{2A}}{D_{1A}} \Omega-\delta \omega.
\end{equation}
The Raman-induced SSFS is a negative frequency shift ($\Omega<0$) with a magnitude that increases with soliton bandwidth and average power. Accordingly, with increasing soliton power (and bandwidth), the plot of the r.h.s. of eqn. (4) versus $\mu$ acquires an increasingly negative slope (green dashed line in fig. 2a). The phase matching mode number, $\mu = \mu_{DW}$, therefore also increases (i.e., the dispersive wave shifts to a higher optical frequency) with soliton power.  The two soliton spectra presented in fig. 2b illustrate this effect (red spectrum is lower power and has the lower dispersive wave frequency). Fig. 2c also shows a series of higher-resoluton scans of the dispersive wave with soliton power increasing from the lower to upper scans and is, again, consistent with the prediction. 

The frequency shift, $\Omega$, repetition frequency, $\omega_r$, and the dispersive wave frequency were measured for a series of soliton powers. The soliton power was varied by changing the cavity-pump detuning frequency ($\delta \omega$) using the method described elsewhere \cite{yi2015soliton,yi2016active}.  $\omega_r$ was measured using an electrical spectrum analyzer after photodetection of the resonator optical output. The offset frequency $\Omega$ was directly measured on an optical spectrum analyzer by fitting the center of optical spectrum (see fig. 1b) to determine the spectral maximum and measuring the wavelength offset relative to the pump. The dispersive wave frequency increases with increasing $\Omega$ (fig. 2d) which is again consistent with the graphical interpretation of eqn. (4). The soliton repetition rate vesus $\Omega$ (see fig. 2d) is fitted using eqn. (3).  The intercept closely agrees with $D_{1A}$ and the slope allows determination of $D_{2A}/2 \pi = 14.7$ kHz (in good agreement with $15.2$ kHz from fitting to the measured dispersion curve in fig. 1a). 

Spatial mode interaction provides a new way to phase match a DKC soliton to a dispersive wave. The approach requires a resonator to support at least two spatial modes. At present, all high-Q systems used for generation of solitons satisfy this condition. It has been shown theoretically and through measurement that the dispersive wave frequency can be actively tuned because of coupling to the soliton offset frequency $\Omega$. In the silica microcavities tested here, this offset is dominated by the Raman-induced SSFS and the dispersive wave is predicted and observed to tune to higher frequencies with increasing soliton power and bandwidth. As a further test of the theory, the dependence of repetition frequency on SSFS was combined with measurement to extract resonator dispersion parameters, which compared well with direct measurements based on resonator dispersion characterization. The dispersion induced by modal interactions in the tested device has been measured and accurately modeled using a coupled-mode formalism. 

Modal interactions such as those studied here for dispersive wave generation can in principle be engineered. This capability has been demonstrated in silica resonators, but with the objective to exclude modal crossings from specific spectral regions so as to enable soliton formation. It seems possible that more complex resonator designs could not only engineer the placement of these crossings, but also locate multiple avoided crossings near a soliton so as to induce multiplets of dispersive waves. 

Note: during submission of this work, Matsko, et. al., reported on Cherenkov radiation by avoided mode crossings in microresonators \cite{matsko2016cherenkov}.

\section*{Funding Information}
The authors gratefully acknowledge the Defense Advanced Research Projects Agency under the PULSE and DODOS programs, NASA, the Kavli Nanoscience Institute.


\noindent 

\bibliography{ref}
\end{document}